\documentclass{appolb}
\usepackage{epsfig}
\usepackage[T1]{fontenc}

\begin{document}
\title{ON SHOCK WAVES IN MODELS\\ WITH V-SHAPED POTENTIALS%
\thanks{Paper supported in part by ESF Programme "\,COSLAB\,".}%
}
\author{Pawe\l \,\,\,Klimas
\address{M. Smoluchowski Institute of Physics, Jagellonian University\\ Reymonta 4, 30-059 Cracow, Poland}
\headauthor{P. Klimas} \headtitle{On Shock Waves in Models with
V-Shaped Potentials}
}
\maketitle
\begin{abstract}
The recently found shock wave solution in the scalar field model
with the field potential $V(\phi)=|\phi|$ is generalized to the
case $V(\phi)=|\phi|-\frac{1}{2}\lambda\phi^2$. We find two kinds
of the shock waves, which are analogous of compression and
expansion waves. The dependence of the waves on the parameter
$\lambda$ is investigated in detail.
\end{abstract}

\section{Introduction}
Interesting and still poorly studied group of field-theoretic
models are these with V-shaped potentials. Physical systems with
just few degrees of freedom and V-shaped potential are studied
quite frequently. There are numerous results for systems such as
{\it e.g.} bouncing oscillators. These systems are mainly studied
in context of chaotical behaviour and grazing bifurcation [1-7].
The V-shaped potentials appear in research of plasma physics [8].
Furthermore, they can play an important role in pinning phenomena
which can describe a process of vortices attaching to lines of
impurities [9,10]. Apart from the applications, they are also very
interesting on purely theoretical grounds because of scale
invariance of vacuum sector, [11].

\indent There are only few analytical results for the
field-theoretic models with V-shaped potentials. An example of
such model, which originally derives from a mechanical system, has
been proposed in [12]. The model considered there is obtained as a
continuum limit of the system of coupled pendulums that are
allowed to take the angular positions belonging to the interval
[$-\phi_0$, $\phi_0$]. This leads to appearance of a V-shaped
potential, and consequently to nontrivial dynamics of the system.
The angle $\phi=0$ corresponds to the pendulum in upward vertical
position. A simpler model has been proposed in [11]. It derives
from coupled system of bouncing balls. The potential in that model
has the form $V(\phi)=|\phi|$. It can be regarded as a limit case
of a large group of V-shaped symmetric potentials.

\indent In the present paper we consider the classical scalar
field model with the potential
$V(\phi)=|\phi|-\frac{1}{2}\lambda\phi^2$, where $\lambda$ is a
real constant. Such potentials with $\lambda\neq 0$ naturally
appear in the case of system of coupled pendulums [12], as well as
in the system of bouncing coupled balls obtained from system
studied in [11] by adding new couplings, see Fig.2. Specifically,
we investigate shock wave solutions.\footnote{In this and previous
papers we use terminology "\,shock wave\," so as to name
discontinuity that moves. There is a criterion in theory of
hydrodynamics that distinguishes between shock waves and different
kinds of discontinuities as e.g contact discontinuity, tangential
discontinuity etc, see e.g. [15]. Our system is clearly different
from e.g. gas medium and the criterion can not be applied
directly. Every choice of name for discontinuity in model
considered here can merely reflect some kind of analogy between
the discontinuity we describe and hydrodynamical discontinuity
that has this name. For this reason we shall remain at the name
"\,shock wave\," and discard looking for a better term.} In the
particular case of $\lambda = 0$ they have been discussed in [13].

\indent The field-theoretic model with the potential $|\phi|$ has
special symmetry. If $\phi(x,t)$ is solution of equation of
motion, then $\nu^2\phi(\frac{x}{\nu},\frac{t}{\nu})$, where $\nu$
is a positive constant, obeys this equation as well, see [13].
This symmetry is "on shell" type because the action functional is
not invariant with respect to the scaling transformation. In
general, in real physical system, apart from term $|\phi|$, the
V-shaped potential has also another terms. A squared term is an
example of the simplest perturbation that breaks the scaling
symmetry. Investigation of effects of such perturbation is a very
important issue.

\indent Our paper is organized as follows. In Sec.2 we show
connections between the model and physical systems. Next section
contains analysis of solutions that have the properties of
expansion shock waves. In Sec.4 we investigate symmetric
compression shock waves. Sec.5 summarizes the paper.

\section{The modified model and related physical systems}

\subsection{General definition}

Let us consider the model of a real scalar field $\phi$ in one
dimension. The Lagrangian has the form
\begin{eqnarray}
L=\frac{1}{2}(\partial_t\phi)^2-\frac{1}{2}(\partial_x\phi)^2-V(\phi),
\end{eqnarray}
where $\phi=\phi(x,t)$ depends on rescaled position $x$ and time
$t$ which are dimensionless. The potential has the form
\begin{eqnarray}
V(\phi)=|\phi|-\frac{1}{2}\lambda\phi^2.
\end{eqnarray}
The evolution equation which corresponds to Lagrangian (1) has the
form
\begin{eqnarray}
(\partial^2_t-\partial^2_x)\phi + \hbox{sign}
(\phi)-\lambda\phi=0.
\end{eqnarray}
Because of sign function equation (3) is clearly nonlinear one. We
assume that $\hbox{sign}(0)=0$. There are three qualitatively
different cases: $\lambda>0$, $\lambda<0$, and $\lambda=0$ when we
have the canonical model considered in [11,13].

\subsection{Positive values of $\lambda$}

Let us start from the case $\lambda>0$. It is convenient to set
$\lambda=\rho^2$. The equation of motion for this case takes the
form
\begin{eqnarray}
(\partial^2_t-\partial^2_x)\phi + \hbox{sign} (\phi)-\rho^2\phi=0.
\end{eqnarray}
\begin{figure}[h]
\centering \leavevmode
\includegraphics[width=0.55\textwidth]{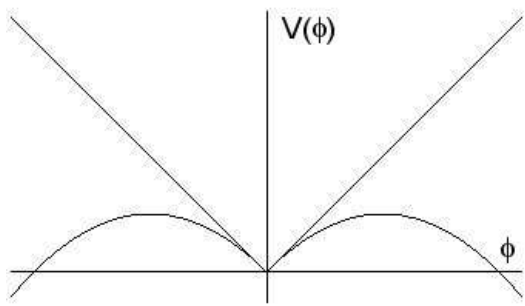}
\caption{The potential $V(\phi)$ ($\lambda>0$) and its limit for
$\lambda=0$.}
\end{figure}
The potential has one local minimum at $\phi=0$ and two local
maxima at $\phi=\pm\frac{1}{\rho^2}$, see Fig.1. It is not
differentiable at its minimum (right-hand side and left-hand side
derivatives are not equal at this point). Equation (4) can be
obtained from equation of motion which describes a small
perturbation around the ground state in the model considered in
[12]. Physical values of perturbation are given by $|\phi|$.

\subsection{Negative values of $\lambda$}

It turns out that the model for $\lambda<0$ has also a physical
meaning. Physical system which is related to this model can be
obtained from the system considered in [11] by adding new springs
that link every ball with the floor, see Fig.2.
\begin{figure}[h!]
\centering \leavevmode
\includegraphics[width=0.6\textwidth]{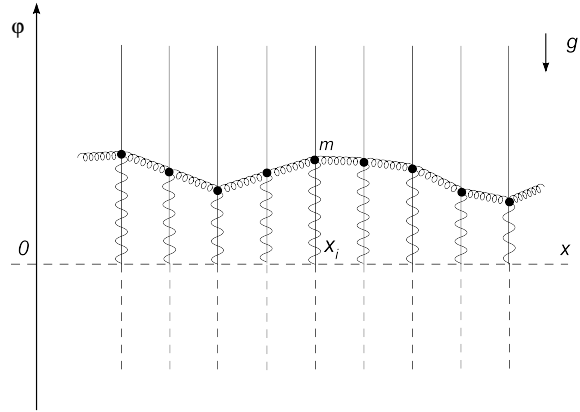}
\caption{The system of bouncing balls.}
\end{figure}
In this system every ball can move only in vertical direction.
There is a rigid floor at $\phi =0$ and every ball bounces
elastically from the floor. After taking few standard steps
(continuous limit, folding transformation) we get the system which
dynamics is described by the equation
\begin{eqnarray}
(\partial^2_t-\partial^2_x)\phi + \hbox{sign}
(\phi)+\sigma^2\phi=0,
\end{eqnarray}
where $-\infty<\phi<\infty$ whereas physical position of the balls
are given by $|\phi|$. As above, we have set $\lambda=-\sigma^2$.
The potential Fig.3 has one minimum and no local maxima.
\begin{figure}[h!]
\centering \leavevmode
\includegraphics[width=0.4\textwidth]{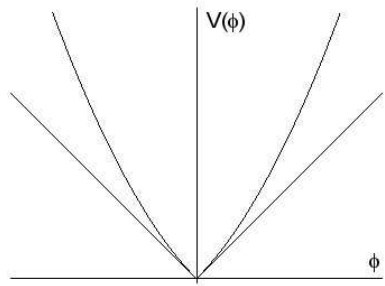}
\caption{The potential $V(\phi)$ ($\lambda<0$) and its limit for
$\lambda=0$.}
\end{figure}

\subsection{On regularized potentials}

\indent In our model at $\phi =0$ first derivative of the
potential does not exist at all. It is possible to replace the
sharp potential by a regularized potential where instead of
$|\phi|$ we use {\it e.g.} $\sqrt{\phi^2+\varepsilon^2}$ or
$\varepsilon \ln{\cosh{\frac{\phi}{\varepsilon}}}$, what gives
well defined first derivative at $\phi =0$. We are not interested
in regularized potential in this paper because physical systems
that we consider here give rather sharp than regularized
potential. Another reason is that for the regularized potential
only solutions such that $|\phi|\gg \varepsilon$ can survive the
limit $\varepsilon \rightarrow 0$. Finding any such solution of
the equation (3) with the regularized potential seems to be a very
difficult task.
\section{Symmetric shock waves inside the light cone}

\subsection{The Ansatz}

\indent Equation (3) includes term with derivatives in the form of
d'Alembert operator so it can be reduced to an ordinary
differential equation by assumption $\phi(x,t)=W(z)$, where
$z=(x^2-t^2)/4$.

\indent There are two qualitatively different cases ($z<0$ and
$z>0$) for the canonical model ($\lambda=0$). When $z<0$ the
pieces of solution can be combined together in one solution. In
opposite case whole solution can be either positive or negative
and it is not limited respectively from above or from below. It is
physically reasonable to get rid of unstable solutions. It can be
simply done by modification our Ansatz to the following one
$\phi(x,t)=\Theta(-z)W(z)$, where $\Theta(-z)$ is well known
Heaviside step function. This modification introduces
discontinuities at $x=\pm t$.

\indent In the modified model, especially for $\lambda>0$ there
are solutions for $z>0$ that are limited both from above and from
below so this time we can not simply get rid of them. In further
part of our work we analyze solutions for $z<0$ and $z>0$
separately because the solution inside the light cone is
completely independent of the solution outside it. It is executed
by assumptions $\phi(x,t)=\Theta(-z)W(z)$ and
$\phi(x,t)=\Theta(z)W(z)$.

\indent Apart from possibility of reduction equation (3) to an
ordinary differential equation, an important question is whether
discontinuities in our model can move with velocity $v\neq 1$ or
not. To answer it let us consider $\phi(x,t)=\Theta(-z)W(z)$,
where $z=pq$, $p=(x-vt)/2$, $q=(x+vt)/2$ and $v\neq 0$. Equation
(3) takes the form
\[
\Theta(-z)\left[A(p,q)W''-\frac{1}{2}(v^2+1)(z
W''+W')+\hbox{sign}(W)-\lambda
W\right]+\nonumber
\]
\[
-2\delta(z)A(p,q)W'(z)-\delta '(z)A(p,q)W(z)=0,\nonumber
\]
where
\[
A(p,q)=\frac{1}{4}(v^2-1)(p^2+q^2),
\]
and $'=\frac{d}{dz}$. The terms proportional to $\Theta(-z)$,
$\delta(z)$ and $\delta'(z)$ have to vanish independently. At
$z=0$ and $v\neq 1$, $W(0)=0$ and $W'(0)=0$ because $A(p,q)\neq
0$. It means that $\phi(x,t)$ can not be discontinuous function
unless $v=1$. We can see that velocity $v=1$ is distinguished by
the model and it is the only admissible velocity with which
discontinuities can move.

\subsection{Equations of motion}

\indent Let us consider the Ansatz
\begin{eqnarray}
\phi(x,t)=\Theta(-z)W(z),\qquad\hbox{where}\qquad
z=\frac{1}{4}(x^2-t^2).
\end{eqnarray}
Applying Ansatz (6) to equation (3) we get the following
differential equation
\begin{eqnarray}
zW''+W'-\hbox{sign}(W)+\lambda W=0.
\end{eqnarray}
Let us introduce a new variable $y$ which is related to $z$ in the
following way
\begin{eqnarray}
z=-\frac{1}{4}y^2.
\end{eqnarray}
Consequently, equation (7) acquires more familiar form
\begin{eqnarray}
G''+\frac{1}{y}G'-\rho^2G=-\hbox{sign}(G),\qquad \lambda>0
\end{eqnarray}
or
\begin{eqnarray}
F''+\frac{1}{y}F'+\sigma^2F=-\hbox{sign}(F),\qquad \lambda<0.
\end{eqnarray}
Equations (9) and (10) are Bessel equations with the signum
nonlinearity. We have denoted $W(-\frac{1}{4}y^2)=G(y)$ for
positive values of $\lambda$ and $W(-\frac{1}{4}y^2)=F(y)$ for
opposite case.

\subsection{Solutions for $\lambda>0$}

The term $\hbox{sign}(G)$ is constant (equal $\pm 1$) on the
intervals where sign of $G(y)$ is constant. Equation (9) has the
solutions:
\begin{eqnarray}
G_{+}(y)&=&\frac{1}{\rho^2}-\alpha I_0(\rho y)-\beta K_0(\rho y)
\,\,\,\,\,\qquad\hbox{for}\qquad G(y)>0,\nonumber\\
G_{-}(y)&=&-\frac{1}{\rho^2}+\widetilde{\alpha} I_0(\rho
y)+\widetilde{\beta} K_0(\rho y) \qquad\hbox{for}\qquad
G(y)<0,\nonumber
\end{eqnarray}
where $\alpha,\beta,\widetilde{\alpha},\widetilde{\beta}$ are
arbitrary constants. We will find these coefficients matching
pieces of solution so as to obtain a solution that is valid on
full available range of $y$. Physically, only real solutions are
interesting so only $y\geq 0$ is considered. It is convenient to
introduce the coefficients $\alpha_k$ and $\beta_k$ and combine
all the solutions in one formula
\begin{eqnarray}
G_k(y)=(-1)^k\left(\frac{1}{\rho^2}-\alpha_kI_0(\rho
y)-\beta_kK_0(\rho y)\right),
\end{eqnarray}
where $k=0,1,2,\ldots$. We assume that $G_0(y)$, that is $G_{+}$
type, starts from $y=0$. It does not cause loosing of generality
because physically relevant quantity is $|\phi|$, thus case where
$G_{-}(y)$ plays role of solution that start from $y=0$ does not
have to be extra discussed. For $G_0(y)$ the coefficient $\beta_0$
has to vanish to ensure regularity at $y=0$. The coefficient
$\alpha_0$ can be expressed with help of $G_0(0)$, and then
$G_0(y)$ takes the form
\begin{eqnarray}
G_0(y)=\frac{1}{\rho^2}-\left(\frac{1}{\rho^2}-
G_0(0)\right)I_0(\rho y).
\end{eqnarray}
For fixed $\rho$ there are three qualitatively different cases
depending on $G_0(0)$.

\begin{figure}[h!]
\centering \leavevmode
\includegraphics[width=0.65\textwidth]{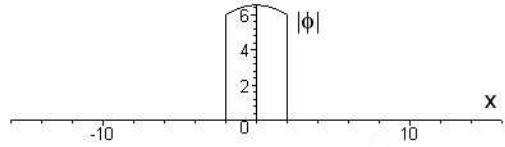}
\caption{The unstable symmetric shock wave at t=2.}
\end{figure}

\begin{figure}[h!]
\centering \leavevmode
\includegraphics[width=0.65\textwidth]{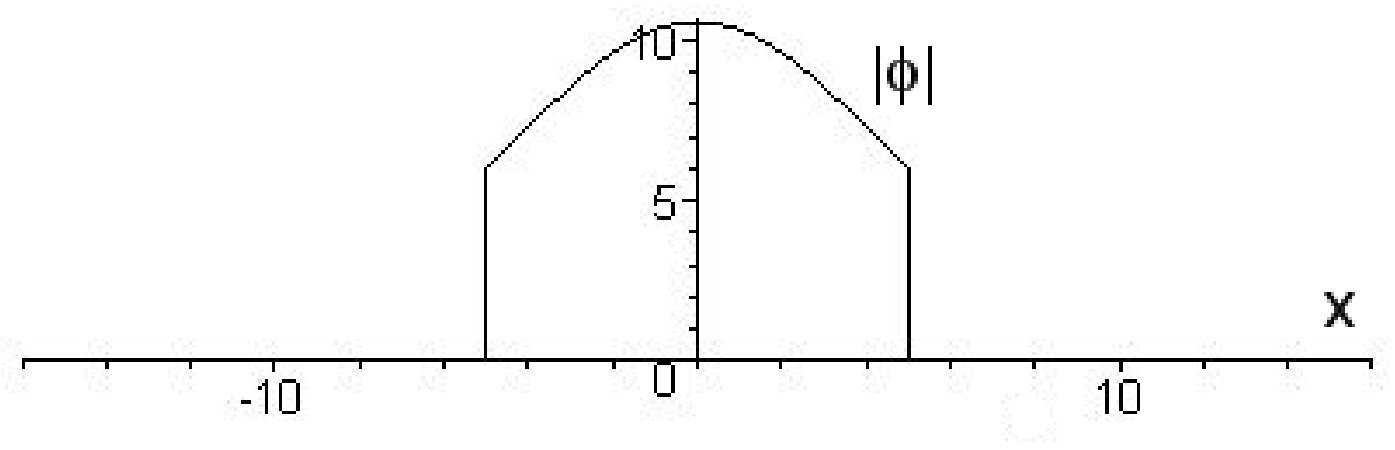}
\caption{The unstable symmetric shock wave at t=5.}
\end{figure}

\begin{figure}[h!]
\centering \leavevmode
\includegraphics[width=0.65\textwidth]{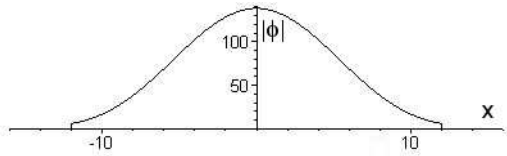}
\caption{The unstable symmetric shock wave at t=12.}
\end{figure}

\indent For $G_0(0)>\frac{1}{\rho^2}$ the solution
$G_0(y)\rightarrow\infty$ for $y\rightarrow\infty$ (unstable
solution). In this case solution $G_0(y)$ cover whole range $y\geq
0$. The shape of the shock wave for $\rho=0.5$ and $G_0(0)=6.0$ at
different times is shown in Figs.4-6.

\indent In the case $G_0(0)=\frac{1}{\rho^2}$ we obtain the shock
wave
\begin{eqnarray}
\phi(x,t)=\frac{1}{\rho^2}\Theta(t^2-x^2).\nonumber
\end{eqnarray}
For this wave, values of the field behind the wave front are
constant and equal $1/\rho^2$.

\indent Last case $G_0(0)<\frac{1}{\rho^2}$ is more complicated
but it is much more interesting. The solution $G_0(y)$ holds only
on the interval $0\leq y<b_0$, where $G_0(y)>0$. For fixed
$G_0(0)$, the first zero of $G(y)$ {\it i.e.} $b_0$ is determined
by solution of following equation
\begin{eqnarray}
I_0(\rho b_0)=\frac{1}{1-G_0(0)\rho^2}.\nonumber
\end{eqnarray}
Unfortunately, it can be solved only numerically. It is clear that
$b_0$ depends on $\rho$. For given $b_0$ first piece of solution
$G(y)$ takes the form
\begin{eqnarray}
G_0(y)=\frac{1}{\rho^2}\left[1-\frac{I_0(\rho y)}{I_0(\rho b_0
)}\right].
\end{eqnarray}
We are interested in a solution for all nonnegative values of $y$.
Having pieces of solution (11) and matching conditions
\begin{eqnarray}
G_k(b_{k-1})=0,\qquad G'_k(b_{k-1})= G'_{k-1}(b_{k-1}),
\end{eqnarray}
which are implied by equation (9) we can calculate coefficients
$\alpha_k$ and $\beta_k$. First matching condition allows to
eliminate coefficients $\beta_k$. Solution (11) takes the form
\begin{eqnarray}
G_k(y)&=&(-1)^k\left[\frac{1}{\rho^2}\left(1-\frac{K_0(\rho
y)}{K_0(\rho b_{k-1})}\right)\right. \nonumber\\
    &-&\alpha_{k}I_0(\rho b_{k-1})\left.\left(\frac{I_0(\rho y)}{I_0(\rho
    b_{k-1})}
    -\frac{K_0(\rho y)}{K_0(\rho b_{k-1})}\right)\right].
\end{eqnarray}
The zeros $b_k$ for $k=1,2,\ldots$ come from the equation
$G_k(b_k)=0$. Like for $k=0$ we can get them numerically. The
second condition in (14) gives coefficients $\alpha_k$:
    \[
    \alpha_1=\frac{1}{\rho^2I_0(\rho b_0)}\frac{\mathcal{K}(b_0,b_0)-\rho^2\alpha_0 I_1(\rho b_0)}
    {\mathcal{I}(b_0,b_0)+\mathcal{K}(b_0,b_0)}
    \]

    \[
    \alpha_k=\frac{1}{\rho^2I_0(\rho b_{k-1})}
    \left[A-\rho^2I_0(\rho b_{k-2})B\right],
    \]
where
    \[
    A\equiv\frac
    {\mathcal{K}(b_{k-1},b_{k-2})+\mathcal{K}(b_{k-1},b_{k-1})}
    {\mathcal{I}(b_{k-1},b_{k-1})+\mathcal{K}(b_{k-1},b_{k-1})},
    \]

    \[
    B\equiv\frac
    {\mathcal{I}(b_{k-1},b_{k-2})+\mathcal{K}(b_{k-1},b_{k-2})}
    {\mathcal{I}(b_{k-1},b_{k-1})+\mathcal{K}(b_{k-1},b_{k-1})}.
    \]
and $k=2,3,\ldots$. We have introduced the special notation

    \[
    \mathcal{I}(x,y)\equiv\frac{I_1(\rho x)}{I_0(\rho y)},\qquad
    \mathcal{K}(x,y)\equiv\frac{K_1(\rho x)}{K_0(\rho y)}.
    \]

\begin{figure}[h!]
\centering \leavevmode
\includegraphics[width=0.7\textwidth]{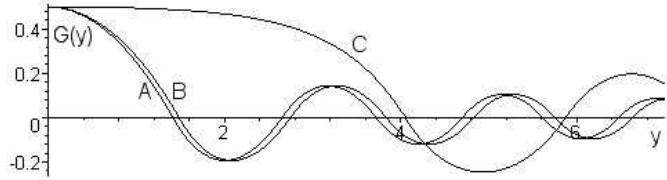}
\caption{The solutions $G(y)$ for fixed $G_0(0)=0.5$ and different
values of $\rho$. The curve $A$ correspond to $\rho=0.1$, $B$ to
$\rho=0.5$ and curve $C$ is the solution for $\rho =1.4$. For
fixed $G_0(0)$ the parameter $\rho$ belongs to the interval
$0<\rho<1/\sqrt{G_0(0)}$.}
\end{figure}

\noindent The zeros of $G(y)$ depend on $\rho$ - they are larger
for larger values of $\rho$, Fig.7. For fixed $G_0(0)$ and
$\rho\rightarrow G_0(0)^{-1/2}$ first zero $b_0\rightarrow\infty$
and, of course, $b_k\rightarrow\infty$ ($b_0<b_1<b_2<\ldots$). All
the zeros $b_k$ are larger than their counterparts $a_k$ in the
canonical model. It means that zeros
$x_k=\pm\left(t^2-b_k^2\right)^{1/2}$ in the modified model run
faster than $x^c_k=\pm\left(t^2-a_k^2\right)^{1/2}$ in the
canonical model. A pair of zeros $x_k$ appears at $t=b_k$ and
moves with velocities $v_k=\pm\left(1-b_k^2/t^2\right)^{-1/2}$. In
Fig.8-10 we present three snapshots which show the symmetric shock
wave for $G_0(0)<\frac{1}{\rho^2}$.

\begin{figure}[!h]
\centering \leavevmode
\includegraphics[width=0.7\textwidth]{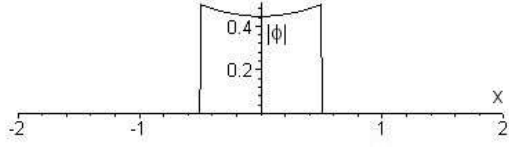}
\caption{The symmetric shock wave for $\rho=0.5$, $G_0(0)=0.5$ at
t=0.5.}
\end{figure}

\begin{figure}[!h]
\centering \leavevmode
\includegraphics[width=0.7\textwidth]{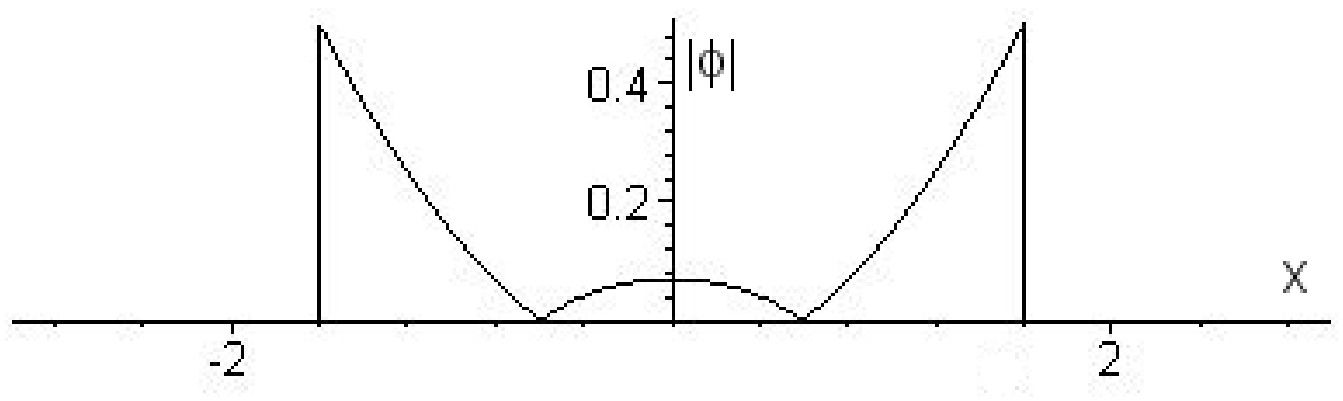}
\caption{The symmetric shock wave for $\rho=0.5$, $G_0(0)=0.5$ at
t=1.6.}
\end{figure}

\begin{figure}[!h]
\centering \leavevmode
\includegraphics[width=0.7\textwidth]{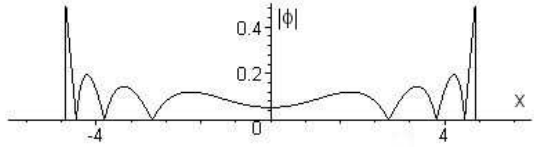}
\caption{The symmetric shock wave for $\rho=0.5$, $G_0(0)=0.5$ at
t=4.7.}
\end{figure}

\subsection{Solutions for $\lambda<0$}

This section is devoted to presentation of solutions correspond to
equation (10). Many steps are the same so we sometimes skip the
comments. Let us start from the solution of (10) given in the form
\begin{eqnarray}
F_k(y)=(-1)^k\left(\mu_kJ_0(\sigma y)+\nu_kY_0(\sigma
y)-\frac{1}{\sigma^2}\right),
\end{eqnarray}
where $J_0$ and $Y_0$ are Bessel functions.  They take real values
for $y\geq 0$. $F_k(y)$ are positive for $k=0,2,4,\ldots$ and
negative for $k=1,3,5,\ldots$. We assume that $F_0(y)$ starts from
$y=0$. This time, for given $\sigma$ there is no qualitatively
change in behaviour of solution for different $F_0(0)$. $F_0(y)$
can be expressed in the form
\begin{eqnarray}
F_0(y)=\left(F_0(0)+\frac{1}{\sigma^2}\right)J_0(\sigma
y)-\frac{1}{\sigma^2}.
\end{eqnarray}
The first zero $c_0$ is calculated from the equation $F_0(c_0)=0$.
We can rewrite $F_0(y)$ using $c_0$
\begin{eqnarray}
    F_0(y)=\frac{1}{\sigma^2}\left[\frac{J_0(\sigma y)}{J_0(\sigma
    c_0)}-1\right].
\end{eqnarray}
In order to have solution for whole range $y\geq 0$ we have to
calculate $\mu_k$ and $\nu_k$. We use the matching conditions
\begin{eqnarray}
F_k(c_{k-1})=0,\qquad F'_k(c_{k-1})= F'_{k-1}(c_{k-1}).
\end{eqnarray}
For $k=1,2,\ldots$
\begin{eqnarray}
    F_k(y)&=&(-1)^{k+1}\left[\frac{1}{\sigma^2}\left(1-\frac{Y_0(\sigma y)}{Y_0(\sigma
    c_{k-1})}\right)-\right.\nonumber \\
    &-&\left.\mu_{k}J_0(\sigma c_{k-1})\left(\frac{J_0(\sigma y)}{J_0(\sigma c_{k-1})}
    -\frac{Y_0(\sigma y)}{Y_0(\sigma c_{k-1})}\right)\right].
\end{eqnarray}
The coefficients $\nu_k$ have been eliminated by using first
condition in (19). The zeros $c_k$ fulfil the equation
$F_k(c_k)=0$. Second condition in (19) gives coefficients $\mu_k$
\[
    \mu_1=-\frac{1}{\sigma^2J_0(\sigma c_0)}\frac{\mathcal{Y}(c_0,c_0)+\mu^2\mu_0 J_1(\rho c_0)}
    {\mathcal{J}(c_0,c_0)-\mathcal{Y}(c_0,c_0)}
\]
\[
    \mu_k=-\frac{1}{\sigma^2J_0(\sigma c_{k-1})}
    \left[\widetilde{A}+\sigma^2J_0(\sigma c_{k-2})\widetilde{B}\right],
\]
where $k=2,3,\ldots$ and
\[
    \widetilde{A}\equiv\frac
    {\mathcal{Y}(c_{k-1},c_{k-2})+\mathcal{Y}(c_{k-1},c_{k-1})}
    {\mathcal{J}(c_{k-1},c_{k-1})-\mathcal{Y}(c_{k-1},c_{k-1})}
\]
\[
    \widetilde{B}\equiv\frac
    {\mathcal{J}(c_{k-1},c_{k-2})-\mathcal{Y}(c_{k-1},c_{k-2})}
    {\mathcal{J}(c_{k-1},c_{k-1})-\mathcal{Y}(c_{k-1},c_{k-1})}.
\]
By analogy, the functions $\mathcal{J}$ and $\mathcal{Y}$ are
defined by formulas
\[
    \mathcal{J}(x,y)\equiv\frac{J_1(\sigma x)}{J_0(\sigma y)},\qquad
    \mathcal{Y}(x,y)\equiv\frac{Y_1(\sigma x)}{Y_0(\sigma y)}.
\]
Having $\mu_k$ we can find consecutive zero $c_k$ as the solution
of the equation $F_k(c_k)=0$. Fig.11 presents solutions $F(y)$ for
$F_0(0)$ that is set and different values of $\sigma$.
\begin{figure}[h!]
\centering \leavevmode
\includegraphics[width=0.8\textwidth]{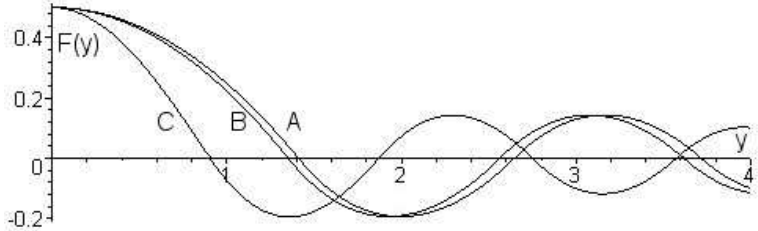}
\caption{The solutions $F(y)$ for fixed $F_0(0)=0.5$ and different
values of $\sigma$. The curve $A$ correspond to $\sigma=0.1$, $B$
to $\sigma=0.5$ and curve $C$ to $\sigma =2.0$. }
\end{figure}
The zeros $c_k$ are smaller than in the canonical model so zeros
$x_k=\pm(1-c_k^2/t^2)^{1/2}$ move slower than zeros $x^c_k$. They
are getting smaller for $\sigma$ being increased. The shape of
$|\phi|$ qualitatively resembles that presented in Fig.8-10.

\subsection{Divergence of sequences $b_k$ and $c_k$}

An interesting problem is if the sequence of $b_k$ (or $c_k$) is
divergent or not. Unfortunately, we can not show a proof of
divergence these sequences (only a numerical evidence) because
explicit expressions for $b_k$ ($c_k$) are not known - we have got
them only as a result of numerical computations. To see it, we
introduce
\[
b_n=b_0x_1x_2\cdot\ldots\cdot x_n,\qquad\hbox{where}\qquad
x_k=\frac{b_k}{b_{k-1}}.
\]
Let us say that for $n\rightarrow\infty$ the sequence of $b_n$ is
divergent $b_n\rightarrow\infty$. It also means that the sequence
of $\ln{b_n}$ is divergent and sum of these logarithms
\[
S_n=\sum_{i=1}^{n}\ln{x_i}
\]
as well. We apply Kummer criterion so as to check if the series is
divergent [16]. In our computation the comparative sequence
$q_n=n\ln{n}$ has been used. It has been checked up to $n=700$
that
\[
\mathcal{K}_n=q_n\frac{b_n}{b_{n+1}}-q_{n+1}
\]
are negative and monotonically decrease, see Fig.12.
\begin{figure}[h!]
\centering \leavevmode
\includegraphics[width=0.7\textwidth]{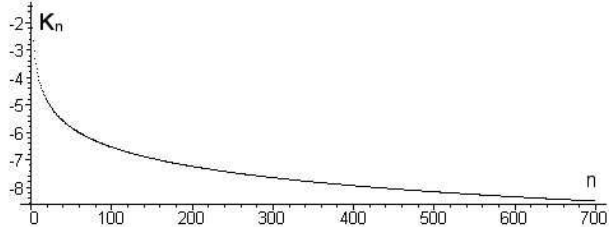}
\caption{Behaviour of $\mathcal{K}_n$ for sequence $b_n$.}
\end{figure}
Kummer criterion says that series is divergent if for all $n>N$,
values of $\mathcal{K}_n\leq0$, where $N$ is a fixed number. It
suggests that solution $G(y)$ covers whole $y\geq 0$. It has been
checked that also the sequence of $c_n$ is probably divergent.

\subsection{The correspondence between the modified and the canonical model}

The potential $V(\phi)=|\phi|$ is the limit case of the modified
one for $\lambda\rightarrow 0$. We are interested in limit
$\lambda\rightarrow 0$ for the solutions in the modified model. It
is not {\it a priori} clear that these limit solutions have to be
solutions that are known from the canonical model. We are able to
do an analytical computation for first two pieces of solution {\it
i.e.} $G_0$ and $G_1$ (or $F_0$ and $F_1$). Let us remind that two
first pieces of solution $W^c(z)$ for the canonical model have the
form
\begin{eqnarray}
W^c_{-1}(z)=z+a_0,\qquad
W^c_0(z)=-\left(z+a_0+2a_0\ln{\frac{|z|}{a_0}}\right),
\end{eqnarray}
where $-a_0$ is the first zero (for variable $z$). For more detail
see [6]. Let us denote $W^{+}_{-1}(z)\equiv G_0(y(z))$ and
$W^{+}_0(z)\equiv G_1(y(z))$ (by analogy $W^{-}_{-1}(z)\equiv
F_0(y(z))$ and $W^{-}_0(z)\equiv F_1(y(z))$). We also rename
$\rho_{+}\equiv\rho$ and $\rho_{-}\equiv\sigma$. We are interested
in comparison of the canonical and the modified solution for
$\rho_{\pm}\rightarrow 0$. In this limit the condition
$W^c_{-1}(0)=W^{\pm}_{-1}(0)$ can be replaced by $a_0=a^{\pm}_0$,
where $a^{+}_0=b^2_0/4$ and $a^{-}_0=c^2_0/4$. The solution
$W_{-1}^{\pm}$ and $W_{0}^{\pm}$ can be expanded in the Taylor
series what gives
\begin{eqnarray}
    W^{\pm}_{-1}(z)=z+a_0\mp C\rho_{\pm}^2+\mathcal{O}(\rho_{\pm}^4),
\end{eqnarray}
\begin{eqnarray}
    W^{\pm}_{0}(z)=-\left(z+a_0+2a_0\ln{\frac{|z|}{a_0}}\right)
    \mp D\rho_{\pm}^2+\mathcal{O}(\rho_{\pm}^4),
\end{eqnarray}
where
\begin{eqnarray}
C\equiv\frac{z^2}{4}
    +a_0z+\frac{3}{4}a_0^2,\nonumber
\end{eqnarray}
\begin{eqnarray}
D\equiv(a_0^2-2a_0z)\ln{\frac{|z|}{a_0}}-\frac{z^2}{4}+3a_0z+\frac{13}{4}a_0^2.\nonumber
\end{eqnarray}
The leading terms in (22) and (23) do not depend on $\rho_{\pm}$
so they are limits of these solutions for $\rho_{\pm}\rightarrow
0$. The most important thing is that these limits are exactly
equal to the first two solutions $W_{-1}^c$ and $W_0^c$ in the
canonical model. The correspondence between the other solutions
$W^{\pm}_k$ can be checked numerically. We can see that the
smaller values of $k$ we take, the better correspondence is.

\section{Symmetric shock waves outside the light cone}

\subsection{Equations of motion}
The solutions outside the light cone can be obtained with help of
the Ansatz
\begin{eqnarray}
\phi(x,t)=\Theta(z)W(z),\qquad\hbox{where}\qquad
z=\frac{1}{4}(x^2-t^2).
\end{eqnarray}
If we now introduce $z=\frac{1}{4}y^2$ we get
\begin{eqnarray}
g''+\frac{1}{y}g'+\rho^2g=\hbox{sign}(g),
\end{eqnarray}
\begin{eqnarray}
f''+\frac{1}{y}f'-\sigma^2f=\hbox{sign}(f),
\end{eqnarray}
where $g(y)=W(z(y))$ for $\lambda>0$ and $f(y)=W(z(y))$ for
$\lambda<0$.

\subsection{Case $\lambda>0$}

It is convenient, for our further analysis of solutions, to
associate the potential $U(g)$ with equation (25). We can easily
see that $U(g)=\frac{1}{2}\rho^2g^2-|g|$, see Fig.13.

\begin{figure}[h!]
\centering \leavevmode
\includegraphics[width=0.6\textwidth]{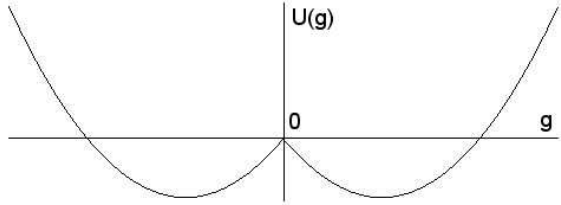}
\caption{The potential $U(g)$.}
\end{figure}

\noindent The solution of (25) takes the form
\begin{eqnarray}
g_k(y)=(-1)^k\left(\frac{1}{\rho^2}-\mu_kJ_0(\rho y)-\nu_kY_0(\rho
y)\right),
\end{eqnarray}
where $g_k>0$ for $k=0,2,4,\ldots$ and $g_k<0$ otherwise. As
above, we have to set $\nu_0=0$ so as to have $g_0(y)$ regular at
$y=0$. If $g_0(0)$ is given then
\begin{eqnarray}
g_0(y)=\frac{1}{\rho^2}-\left(\frac{1}{\rho^2}-g_0(0)\right)J_0(\rho
y).
\end{eqnarray}
In Fig.14 we present a few curves (28) for different values of
$g_0(0)$. There are several qualitatively different cases.
\begin{figure}[h!]
\centering \leavevmode
\includegraphics[width=0.8\textwidth]{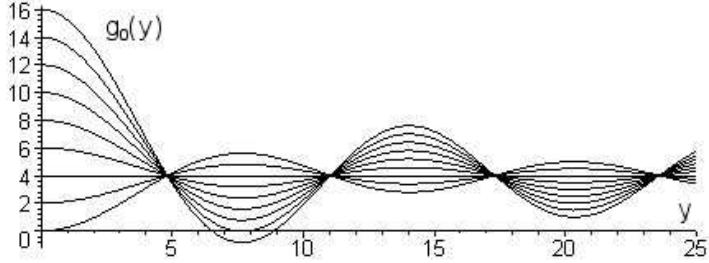}
\caption{The curves $g_0(y)$ for different $g_0(0)$ and
$\rho=0.5$. They are solution for full range $y\geq 0$ except
those which start from $g_0(0)=14$ and $g_0(0)=16$.}
\end{figure}

Case $g_0(0)=0$ corresponds to solution that starts from the point
where $U(g)$ has its local maximum. We have chosen positive
solution but for $g_0(0)=0$, negative solution is possible as
well. We can get these solutions by replacing $(-1)^k$ in (27)
with $(-1)^{k+1}$ and assuming that $g_0(y)$ starts from $y=0$. We
will not discuss this situation separately because $|\phi|$ is
physical quantity. It is worth mentioning that this solution is
not exactly shock type because there is not sharp front wave.

For $0<g_0(0)<g_0^{crit}$, where
\[
g_0^{crit}=\frac{1}{\rho^2}\left(1-\frac{1}{J_0(j^1_1)}\right)
\]
the solutions $g_0(y)$ are valid for all $y\geq 0$. $j^1_1$ is the
first zero of $J_1$, ($j^0_1=0$). Approximately,
$\rho^2g^{crit}_0=3.482872$. This case contains the constant
solution $g_0(y)=\frac{1}{\rho^2}$.

For critical value of $g_0(0)=g_0^{crit}$ the first zero of $g(y)$
appears. The solution can take either of forms: the solution

\begin{eqnarray}
g_0(y)=\frac{1}{\rho^2}\left(1-\frac{J_0(\rho
y)}{J_0(j^1_1)}\right)
\end{eqnarray}
for all $y\geq 0$ or $g_0(y)$ given by (29) for $0\leq y\leq
j^1_1/\rho$ and $g_1(y)=-g_0(y)$ for $y\geq j^1_1/\rho$.

If $g_0(y)$ is a little bit bigger than $g^{crit}_0$ the solution
is made up of $g_0(y)$ and $g_1(y)$ but this time $j^1_1$ in (29)
is replaced with $\rho c_0$ and $g_1(y)$ contains also function
$Y_0(\rho y)$, see Fig.15. This picture is valid until $g_0(0)$
reach consecutive critical value $g_0^{crit2}$ (unfortunately, we
are not able to give appropriate analytical formula - only
numerical value of $g^{crit2}$ is available).

Therefore, for $g_0(0)=g_0^{crit2}$ situation resembles that for
$g_0(0)=g_0^{crit}$ but this time two zeros $c_0$ and $c_1$
already exist.
\begin{figure}[h!]
\centering \leavevmode
\includegraphics[width=0.75\textwidth]{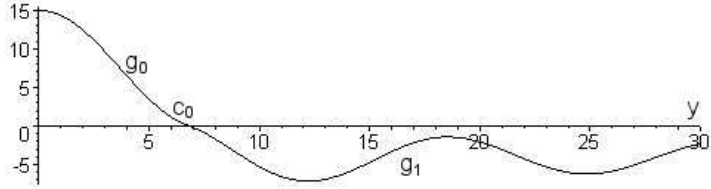}
\caption{The solution $g(y)$ that has only one zero.}
\end{figure}

\begin{figure}[h!]
\centering \leavevmode
\includegraphics[width=0.75\textwidth]{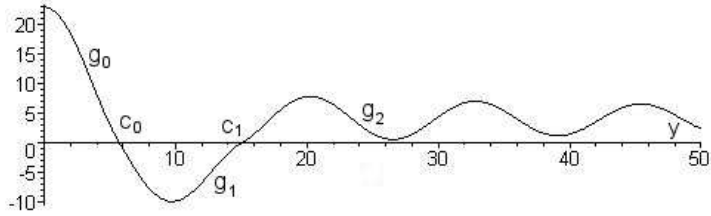}
\caption{The solution $g(y)$ with two zeros.}
\end{figure}

\begin{figure}[h!]
\centering \leavevmode
\includegraphics[width=0.75\textwidth]{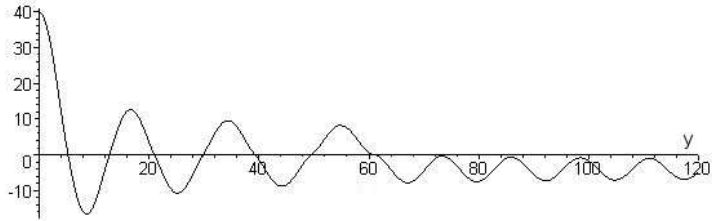}
\caption{The solution $g(y)$ with six zeros.}
\end{figure}

For values of $g_0(0)$ a little bit more than $g_0^{crit2}$
solution looks like it was shown in Fig.16. For greater values of
$g_0(0)$ more zeros $c_k$ appear, see Fig.17. An important issue
is that for finite $g_0(0)$, maximal value of $k$ is always a
finite number. For these solutions, asymptotic values of $g(y)$
for $y\rightarrow \infty$ {\it i.e.} $\pm 1/\rho^2$ correspond to
the minima of the potential $U(g)$. We have not discussed yet the
formulas for coefficients $\mu_k$ and $\nu_k$ in (27).
Fortunately, they have the same form as analogical coefficients in
equation (16) - only $\sigma$ has to be replaced with $\rho$.

\subsection{Case $\lambda<0$}

The potential $U(f)$ for equation (26) takes the form
$U(f)=-\frac{1}{2}\sigma^2f^2-|f|$ what suggests unstable
behaviour of $f(y)$. The solutions of (26) can be written down in
the form
\begin{eqnarray}
f_{+}(y)&=&\alpha I_0(\rho y)+\beta K_0(\rho y)-\frac{1}{\sigma^2}
\,\,\,\,\,\qquad\hbox{for}\qquad f(y)>0,\nonumber\\
f_{-}(y)&=&-\widetilde{\alpha} I_0(\rho y)-\widetilde{\beta}
K_0(\rho y)+\frac{1}{\sigma^2} \qquad\hbox{for}\qquad
f(y)<0.\nonumber
\end{eqnarray}
Let us consider $f(y)>0$. By analogy to our previous analysis we
will denote it as $f_0(y)$. For given $f_0(0)$ it takes the form
\begin{eqnarray}
f_0(y)=\left(\frac{1}{\sigma^2}+f_0(0)\right)I_0(\sigma
y)-\frac{1}{\sigma^2},
\end{eqnarray}
where $\beta=0$. There are five such solutions in Fig.18. The
solution $f_0(y)$ covers whole range $y\geq 0$. We have an
analogical situation for a negative solution. The border solution
($f_0(0)=0$) is, of course, a non-shock type.
\begin{figure}[h!]
\centering \leavevmode
\includegraphics[width=0.75\textwidth]{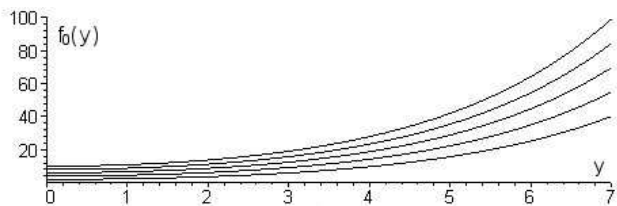}
\caption{The solutions $f_0(y)$ for different (positive) values of
$f_0(0)$.}
\end{figure}

\section{Summary}

We have presented the shock wave solutions in the model with the
potential $V(\phi)=|\phi|-\frac{1}{2}\lambda\phi^2$, where
$\lambda$ is a real constant. The square term plays the role of
perturbation of the potential $V(\phi)=|\phi|$. The potential
$|\phi|$ plays a special role because the corresponding equation
of motion has the scaling symmetry. The square term is the
simplest one that breaks this symmetry. It has been shown that
both cases with nonzero values of $\lambda$ ($\lambda>0$ and
$\lambda<0$) have physical applications - appropriate potentials
appear in the system of coupled bouncing pendulums or bouncing
balls.

\indent Two kinds of waves have been found. The first one is
exactly zero outside the light cone and has two wavefronts (the
field is non-continuous at them) exactly at the surface of the
light cone. Depending on $G_0(0)$ the solution inside the light
cone is unstable, constant, or has isolated zeros. We have found
that these zeros run faster ($\lambda>0$) or slower ($\lambda<0$)
than their counterpart in the canonical model ($\lambda=0$). It
has been also argued (by showing the numerical evidence) that
zeros of solution, that depends on variable $y$, form probably
divergent sequence. Moreover, we have shown (analytically in the
case of first two pieces of solution) that shock waves inside the
light cone for the modified model in the limit $\lambda\rightarrow
0$ reduce to solutions known from the canonical model.

\indent The second type of solution has also wavefronts at $x=\pm
t$. This solution takes zero values inside the light cone. There
is family of solutions that asymptotically ($y\rightarrow\infty$)
reach values $\pm 1/\rho^2$. At the points $g=\pm 1/\rho^2$ the
potential $U(g)$ (Fig.13) has the local minima (or $V(\phi)$ has
its local maxima, Fig.1). Among solutions that belong to this
family, there are solutions that have no zeros, have only one
zero, exactly two zeros {\it etc.} In contrast to the shock waves
inside the light cone that have infinite number of zeros,
solutions considered here have always finite number of zeros.
There is also non-shock type solution - it has only one zero at
the surface of light cone. It can be regarded as the border case
of solutions belonging to this family. We have also found another
family of solutions that have no zeros and grow to infinity for
$y\rightarrow 0$.

\section{Acknowledgements}
I would like to thank Henryk Arod\'z for discussion and numerous
valuable remarks.

\end{document}